\documentclass[12pt]{iopart}

\usepackage{iopams}  
\usepackage{graphicx}
\begin{document}

\comment{Comment on ``Field-driven phase transitions in a
quasi-two-dimensional quantum antiferromagnet"}

\author{A. Zheludev and D. H{\"u}vonen}

\address{Neutron Scattering and Magnetism, Laboratory
for Solid State Physics, ETH Zurich, Switzerland}
\ead{zhelud@ethz.ch}
\begin{abstract}
In a recent publication [M.~B. Stone \etal, New Journal of Physics {\textbf 9} 31 (2007)] a ``Renormalized Classical 2D'' (RC) phase has been reported in a quasi-two-dimensional quantum antiferromagnet PHCC.
Its key signature is a sharp cusp-like feature in the magnetic susceptibility which 
appears below the critical field of magnetic ordering indicated by specific heat anomaly and emergence of a Bragg peak.
Here we present experimental data which shows that regardless of experimental geometry, the specific heat and susceptibility anomalies in PHCC both coincide with the onset of true long range order. This leaves no room for any additional intermediate ``RC'' phase.
\end{abstract}


\submitto{\NJP}
\maketitle

 One of the central claims of this work is the existence of a novel ``Renormalized Classical 2D'' (RC) phase in the quantum magnet PHCC. This phase supposedly separates the long range order (LRO) phase from the quantum paramagnetic (QP) phase. Its key signature is a sharp cusp-like feature in the magnetic susceptibility. According to Stone {\it et al.}, at $T>0$ it appears {\it below} the critical field of magnetic ordering. The latter  manifests itself in a lambda-anomaly of $C(T)$ or $C(H)$, and the appearance of a magnetic Bragg reflections. The RC phase is then said to exist in the field range between the two {\it distinct} anomalies in susceptibility and specific heat.

\begin{figure}
\begin{center}
\includegraphics[width=12cm]{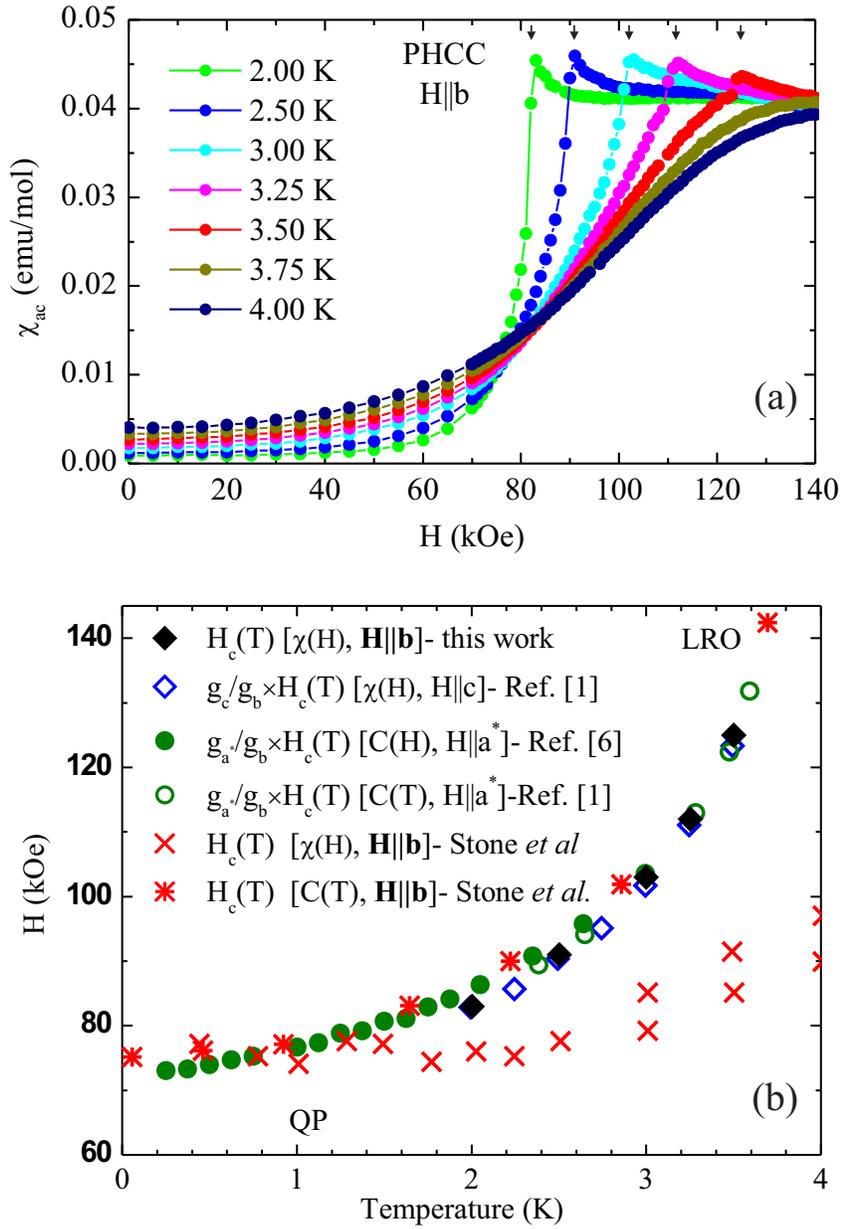}
\caption{ a) Magnetic susceptibility of PHCC measured in $H||b$ orientation. Arrows indicate the susceptibility kinks signifying phase transition. b) $H-T$ phase boundaries in PHCC based on calorimetric and magnetometric data from different sources. 
\label{fig1}}
\end{center}
\end{figure}

In a recent study \cite{Huevonen2013} we have demonstrated that
any RC-like phase is absent for a magnetic field applied along the $c$ crystallographic axis. It was later argued \cite{Broholm2013} that the discrepancy is only apparent, since the RC phase was initially reported for the $H||b$ geometry, and may be absent in other orientations. To clarify this issue we repeated our measurements in the $H||b$ geometry of the original study. Measurements were done on a 69~mg solution-grown \cite{Yankova2012} single crystal of PHCC. AC susceptibility data were collected  at 316~Hz using a commercial Quantum Design PPMS system. The measured $\chi(H)$ curves are shown in Fig.~\ref{fig1}a for different temperatures. As in all previous studies, including those represented in Fig.~2 of Stone et al. in \cite{Stone2007} and Fig.~2 of \cite{Stone2006},  we see a clear cusp-like feature in the measured field sweeps (arrows). The data at each temperature were analyzed as in \cite{Huevonen2013} to pinpoint the exact position of the anomaly. In Fig.\ref{fig1}b we show the measured temperature dependence (solid diamonds). For comparison, we also show the critical field of 3D ordering determined from $C(T)$ measurements, as digitized from Fig.~11 of Stone {\it et al} (asterisks). Within experimental error, the two phase boundaries coincide, excluding the possibility of any intermediate phase. In contrast, the locations of the sharp anomalies in $\chi(T)$ and $\chi(H)$ observed by Stone {\it et al.} are shown in crosses (digitized from Fig.~11), and are clearly inconsistent with our measurements.

The actual field orientation does not seem to play any role, despite \cite{Broholm2013}. In Fig.\ref{fig1}b we show the temperature dependencies of calorimetric and $\chi(T)$ anomalies previously measured in PHCC for different field orientations. For these data, the values of magnetic field were re-normalized by the slightly anisotropic $g$-factor of Cu$^{2+}$ in PHCC \cite{Glazkov2012}. An excellent data collapse is apparent.

A further discrepancy with Stone {\it et al.} is in the shape of the measured $\chi(T)$ curves at elevated temperatures. Our $H||b$ $\chi(H)$ data measured at $T=4$~K  (Fig.\ref{fig1}a) are considerably broader than the corresponding curves in Fig.~2 of Stone {\it et al.}. A similar discrepancy is found between our 3~K data and those reported by Stone {\it et al.} in \cite{Stone2006}. Note that $\chi(T)$ curves from Stone {\it et al.} closely resemble our data collected at lower temperatures. One can speculate that the discrepancy may amount to a trivial thermometry problem. To ensure that this does not apply to our new measurements, we have performed experiments in a different PPMS unit using a vibrating sample magnetometer equipped with separate temperature sensor. The  susceptibility curves obtained by differentiating the thus-measured magnetization data are virtually identical to those that we get from AC experiments.

In summary, regardless of experimental geometry, the specific heat and susceptibility anomalies in PHCC both coincide with the onset of true long range order. This leaves no room for any additional intermediate ``RC'' phase.

\section*{Acknowledgements}
This work is partially supported by the Swiss National Fund under project 2-77060-11 and through Project 6 of MANEP.


\end{document}